\documentclass[a4paper,10pt]{article}
\usepackage[utf8]{inputenc}
\usepackage{amsmath}
\usepackage{amssymb}
\usepackage{graphicx}
\usepackage[colorlinks=false,
            linkcolor=false,
            urlcolor=blue,
            citecolor=false]{hyperref}
\usepackage{graphicx,wrapfig,lipsum}
\usepackage[font=small,belowskip=-5pt,aboveskip=0pt]{caption}
\usepackage{textcomp}
\usepackage{float}
\usepackage[margin=0.5in]{geometry}

\usepackage{url}

\usepackage{caption}
\usepackage{subcaption}

\usepackage{booktabs}
\newcommand{\ra}[1]{\renewcommand{\arraystretch}{#1}}

\DeclareMathOperator*{\argmin}{\arg\!\min}
\title{\textbf{Introducing the ``opacity'' for IMPT planning:\\
       can it improve robustness and quality of plans?}}
\author{F. Arcadu}
\date{February 16, 2015}

\begin{document}

\maketitle


\begin{abstract}
Intensity modulated proton therapy (IMPT) provides highly conformal dose distributions
through the application of multiple, angularly spaced fields, each applying an ad-hoc pattern of spatially varying particle
fluences. In particular, Bragg peaks are simultaneously optimized for all the fields.
Once the number and the direction of the fields are set, the dose distribution of the IMPT plan is determined by the dose
constraints assigned to specific organs-at-risk (OARs) surrounding the tumour.
\newline In this work, we introduce a new feature for the OARs, called \emph{opacity}, aimed at improving the quality and
the robustness of IMPT plans, by modifying, in the pre-optimization stage, the fluence of the pencil beams,
that intersect those structures. The proposed IMPT planning strategy is addressed to those clinical cases, where metallic
prothesis or fast density-varying organs can compromise the stability of the dose distribution. 
We show how the usage of this additional parameter
can lead to more accurate IMPT plans with respect to the situation in which only dose constraints are adopted. 
\end{abstract}

\section{Proposed method}

\subsection{Pencil beam based treatment}
Treatment plans usually involve more the one proton field in order to guarantee 
dose homogeneity over the target volume and to increase the stability of the plan.
The active scanning technique \cite{Pedroni1995} allows, in particular, to control the dose
deposition of each focused pencil beam within the patient volume; this implies that a field is conceived
as the collection of all the single dose spots delivered for a specific position of the patient with respect to the nozzle.
Thank to this peculiarity, the active scanning enables the construction of complex-shaped dose distributions.
Once the number and direction of the proton fields are selected for a treatment, there are two different
planning strategies implemented for active scanning based gantries: \emph{single field uniform dose (SFUD)} \cite{Scheib1993}, implying 
a superposition of the individually optimized dose distributions; \emph{intensity modulated proton therapy
(IMPT)} \cite{Lomax1999}, consisting in a simultaneous optimization of all the fields. The latter modality can often provide better
tradeoffs between the dose coverage of the target and the sparing of the organs at risk.

\subsection{IMPT optimization}
The IMPT optimization consists in the following problem \cite{Albertini2011}:
\begin{equation}
  \boldsymbol{\omega} = \argmin_{\boldsymbol{\omega}} F(\boldsymbol{\omega})
\end{equation}

\begin{equation}
  F(\boldsymbol{\omega}) = \sum\limits_{i_{t}}^{}\left(P_{i_{t}}-D_{i_{t}}(\boldsymbol{\omega})\right)^{2} + 
                           \sum\limits_{i_{o}}^{}g_{i_{o}}^{2}\:\left(P_{i_{o}}-D_{i_{o}}(\boldsymbol{\omega})\right)^{2}
  \label{impt-functional}
\end{equation}
where the first sum runs on the voxel indices of the target $\{i_{t}\}$ and the second one on the voxel indices of the OARs
$\{i_{o}\}$, $P_{j}$ and $D_{j}$ are respectively the prescribed and the computed dose  for the $j$-th voxel, $\{g_{i_{o}}\}$'s are
weights chosen by the planner and $\boldsymbol{\omega} = \{\omega_{1},\cdots,\omega_{N}\}$ corresponds to the fluences
of the pencil beams. The iterative minimizer of (\ref{impt-functional}) is with the OAR dose constraints,
applied through the weights $\{g_{i_{o}}\}$ and with $\boldsymbol{\omega}^{(0)}$, such that each field is characterized by a flat
spread-out bragg peak (SOBP) \cite{Bortfeld1996} within the target volume. Since 
the number and direction of the fields have been already fixed and the pre-optimized fluences $\boldsymbol{\omega}^{(0)}$
are automatically computed on the basis of the target volume, the dose constraints are the only leverage left at disposal of the
planner to modify the resulting dose distribution.

\subsection{Opacity concept}
\begin{figure*}[!h]
  \centering
  \includegraphics[width=6.0in]{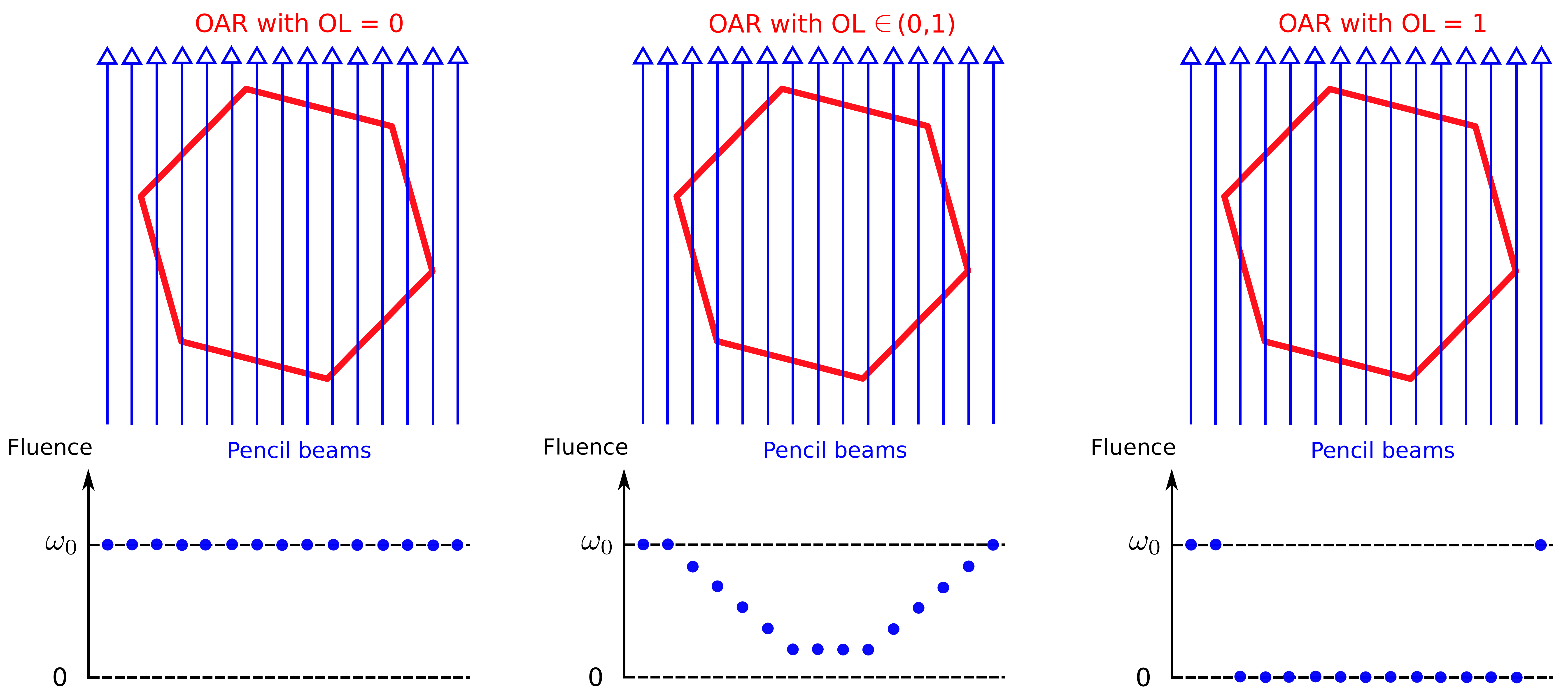}
  \caption{Graphical explanation of the opacity penalization, assuming the pencil beams all characterized by 
           the same initial fluence $\omega_{0}$; on the left, the case of the OAR being ``transparent''; 
           in the middle, the case of the OAR being partially ``opaque'' and penalizing more the pencil beams
           travelling more inside the OAR itself; on the right, case of the OAR completely ``opaque''.}
  \label{opacity-levels}
\end{figure*}

The \emph{opacity} is defined as a quality of the OARs in relation to the pencil beams and has to be considered as a simple weight
ranging in [0,1]. The opacity level (OL) of an OAR is thought to modify the pre-optimized fluences $\boldsymbol{\omega}^{(0)}$
of the pencil beams, whose trajectory
(the straight line connecting the Bragg peak position to the leaving point at the nozzle) intersects voxels of such structure.
We indicate with $\{k_{n}\} \subseteq \{1,\cdots,N\}$ the list of indices referring to the subset of pencil beams crossing the $n$-th
OAR.  
If the OL = 0, the OAR is considered to be ``transparent'' and the pencil beams intersecting the structure are left unaltered,
$\{\tilde{\omega}_{k_{n}}\} = \{\omega_{k_{n}}\}$;
if the OL = 1, the OAR is completely ``opaque'' and $\{\tilde{\omega}^{(0)}_{k_{n}}\} = 0$; when OL $ = \alpha \in (0,1)$ 
is ``partially opaque'' and the pencil beam fluences are penalized in the following way:
\begin{equation}
  \tilde{\omega}^{(0)}_{\bar{k}_{n}} = \alpha \cdot l_{\bar{k}_{n}} \cdot \omega^{(0)}_{\bar{k}_{n}}
  \hspace*{0.6cm} \forall \bar{k}_{n} \in \{k_{n}\} \hspace*{0.3cm},
  \label{smooth-penalization}
\end{equation}
where $l_{\bar{k}_{n}}$ corresponds to the path length travelled by the $\bar{k}_{n}$-th pencil beam
inside the $n$-th OAR. Clearly, in this latter case, the longer the path length inside the OAR, the more the initial fluence
of the pencil beam will be decreased. The three cases are shown in Fig.\ref{opacity-levels}.
\newline
The OL's represent essentially an additional degree of freedom at disposal of the planner to change the starting conditions
and, therefore, to steer the outcome of the IMPT optimization. The dose constraints modify $\boldsymbol{\omega}^{(0)}$ according to the dose 
released by the pencil beam to the OAR, whereas the OL's acts on the basis of the pencil beam trajectory in relation
to the OAR. 
\newline
The OL's penalization requires the knowledge of all the voxel indices that are crossed by every pencil beam.
This computation is efficiently performed by an optimized version of the Siddon algorithm \cite{Jacobs1998}, commonly
adopted for the fast calculation of radiological paths.

\section{Materials}
IMPT plans of different indications and tumour sites have been re-optimized using the OL penalization
and the resulting plans were compared to the delivered ones. In the following, we offer a description of
the clinical cases that were selected for this simulation and of the analysis used to benchmark the robustness
of the dose distributions.

\subsection{Patient with metallic cage}
\begin{figure*}[!h]
  \centering
  \includegraphics[width=3.5in]{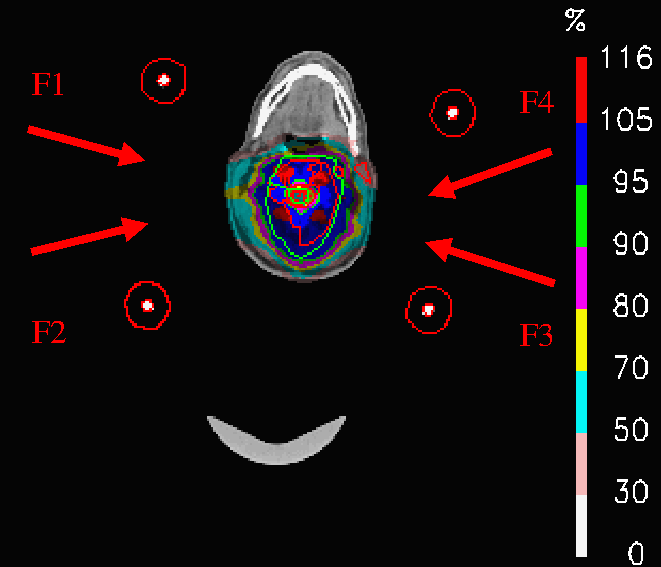}
  \caption{Patient with metallic cage: nominal IMPT dose distribution; inside the patient volume, the green
           contour corresponds to the PTV and the red ones to the OARs; otuside the patient volume,
           the safety VOIs and the effective countours of the 4 metallic rods are visible;
           the fields F1, F2, F3, F4 used for PLAN-NOM and PLAN-OL-1 are also shown.}
  \label{plan-metallic-cage}
\end{figure*}

The first clinical case concerns a patient with an external metallic cage. Patients with neck tumours may undego surgery 
before any radiotherapic treatment with protons and a cage is, therefore, required, to stabilize the head to the neck.
Since no calibration curve of the cage material is available for dose calculation, no pencil beam has to cross the metallic
structure to prevent any wrong placement of Bragg peak inside the patient volume.
\newline
The strategy followed in the nominal treatment
(PLAN-NOM) was to draw safety volumes around the metallic rods and to apply hard dose constraint to such volumes of interest (VOIs),
as shown in Fig.\ref{plan-metallic-cage}, such that the pencil beams of the four fields (F1, F2, F3, F4) involved
in the treatment were safe from hitting the rods.
\newline
The plan has been re-calculated by replacing the safety VOIs and related dose constraints with the effective contours of the rods,
set with OL = 1. In a first attempt, the direction of the fields were kept the same (PLAN-OL-1) and , then, an other plan was
generated by considering a wider angular spacing between F1-F2 and F3-F4 (PLAN-OL-2). The angular increase of the new fields
with respect to the nominal ones was of 10\textdegree.

\subsection{Fields crossing nasal cavities} 
\begin{figure*}[!h]
        \centering
        \begin{subfigure}[b]{0.3\textwidth}
                \includegraphics[scale=0.25]{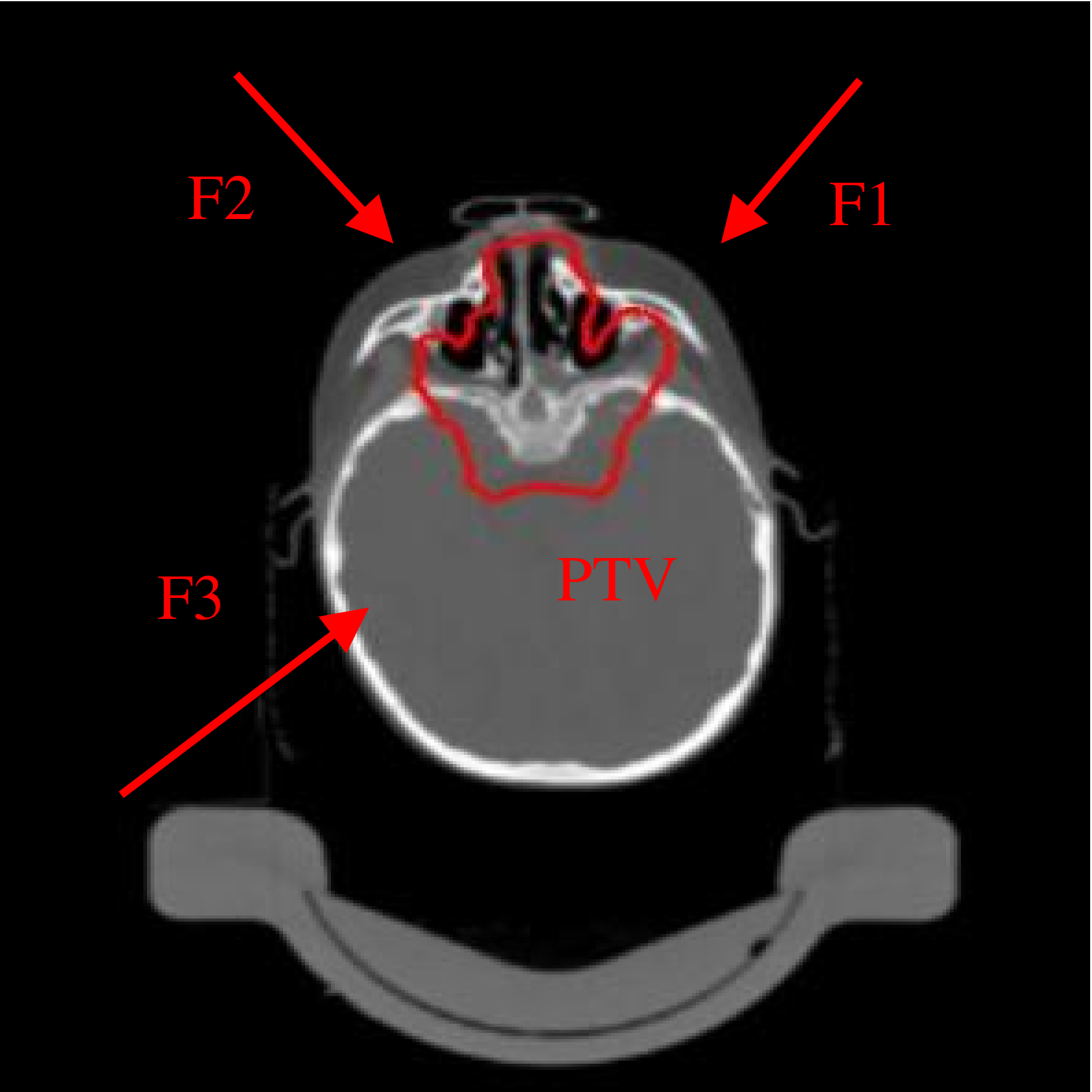}
                \caption{Axial CT slice}
        \end{subfigure}
        \begin{subfigure}[b]{0.3\textwidth}
                \includegraphics[scale=0.25]{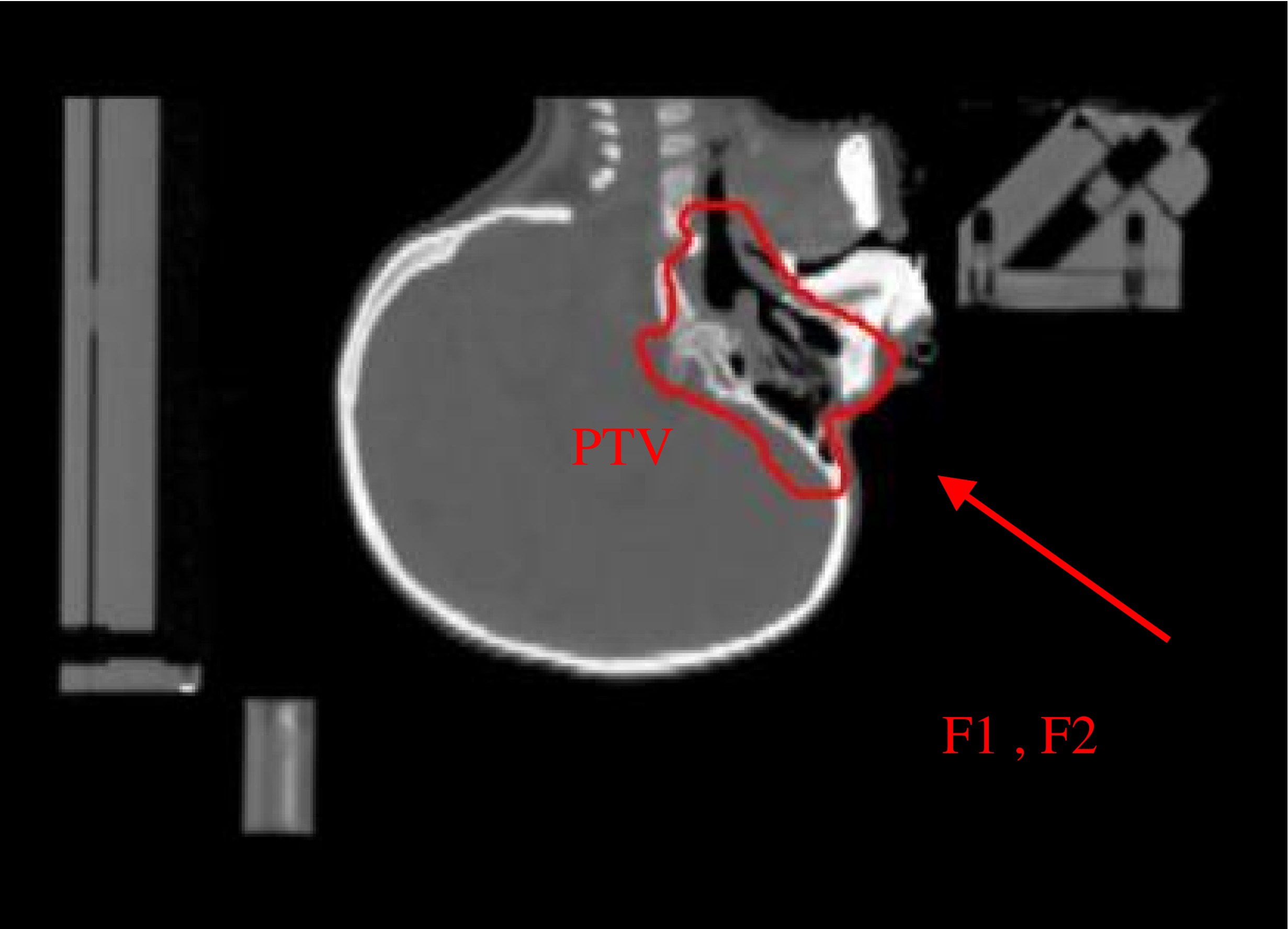}
                \caption{Azimuthal CT slice}
        \end{subfigure}
        \caption{Fields crossing the nasal cavities: axial and azimuthal CT slice of the nasal cavity plan showing the red contour
                 of the PTV; the fields F1, F2 and F3 involved in the treatment are also shown.}
        \label{field-nasal-cavity-plan}
\end{figure*}

\begin{figure*}[!h]
        \centering
        \hspace*{-1cm}
        \begin{subfigure}[b]{0.3\textwidth}
                \includegraphics[scale=0.3]{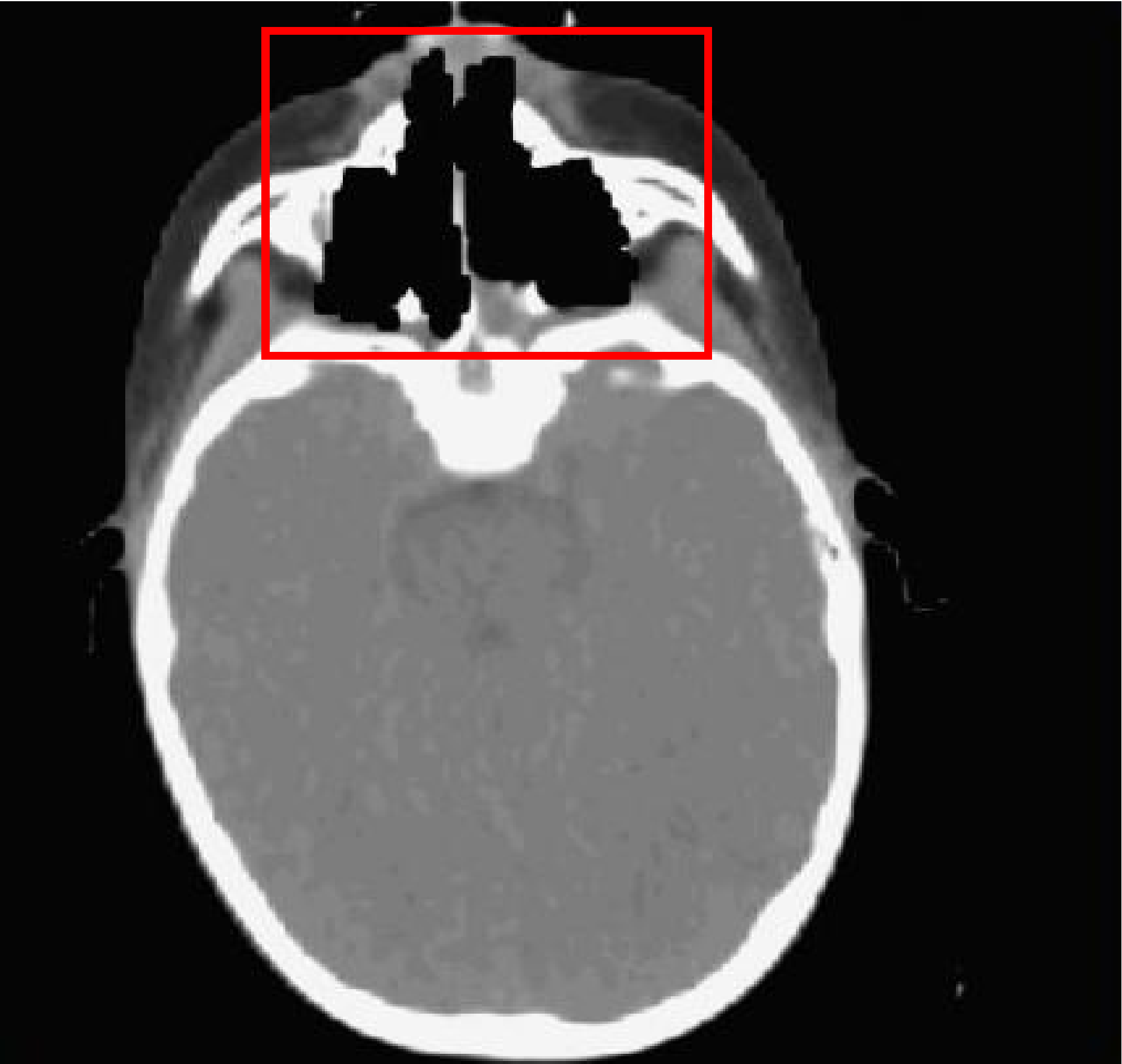}
                \caption{Empty nasal cavity}
        \end{subfigure}
        \hspace*{1cm}
        \begin{subfigure}[b]{0.3\textwidth}
                \includegraphics[scale=0.3]{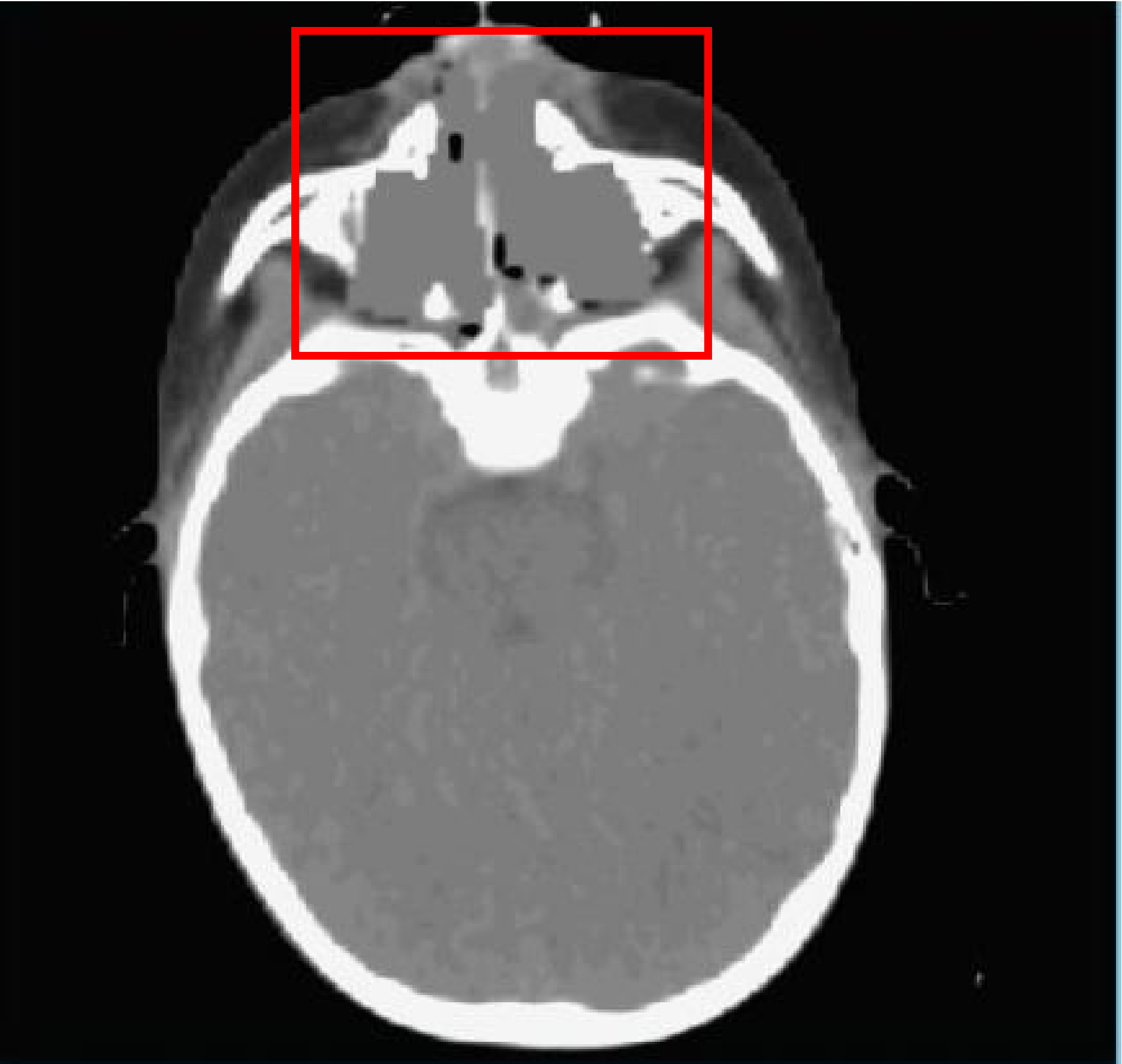}
                \caption{Nasal cavity filled with mucus}
        \end{subfigure}
        \caption{Fields crossing the nasal cavities: modified CT slices simulating th extreme scenarios, when the nasal cavity
                 is completely empty (HU=0) and filled with mucus (HU=30).}
        \label{nasal-cavities}
\end{figure*}

In some clinical cases, the optimal proton fields planned for a treatment cross
anatomical cavities to deliver dose to the target volume. Cavities, like the nose or the bowel, may undergo relevant
density changes between the day of the treatment and the acquisition of the CT image (used to calculate the 
dose distributions for all the treatment fractions). Since the pencil beams crossing the cavity encounter a different
density object, they may be affected by not negligible range uncertanties; in particular, they will undershoot, in case
of a density increase, and overshoot, in the opposite situation.
\newline
The case of a treatment with fields crossing the nasal cavity has been taken into consideration. The field directions 
of the nominal plan are shown in Fig.\ref{field-nasal-cavity-plan}; the frontal fields F1 and F2 encounter
the nasal cavity to reach the target volume.
To evaluate the stability of the IMPT plan, first, the original CT image was modified to simulate the extreme scenarios,
of the cavity being completely empty (HU = 0) and filled with mucus (HU $\simeq$ 30), as shown in Fig.\ref{nasal-cavities};
the original plan (PLAN-NOM) and the plan with OL penalization (PLAN-OL), where an OL $\in (0,1)$ was selected for the nose VOI,
were re-computed for both extreme scenarios, to evaluate the potential variation range with respect to the planning CT. 
We named PLAN-NOM-H and PLAN-NOM-L, respectively, the difference between the nominal plan recomputed on the CT
with low and high density nasal cavity and the plan computed on the original CT 
and we did similarly for the extreme scenario plans with OL penalization (PLAN-OL-H and PLAN-OL-L).

\subsection{Head and neck tumour}
\begin{figure*}[!h]
        \centering
        \hspace*{-1cm}
        \begin{subfigure}[b]{0.3\textwidth}
                \includegraphics[scale=0.3]{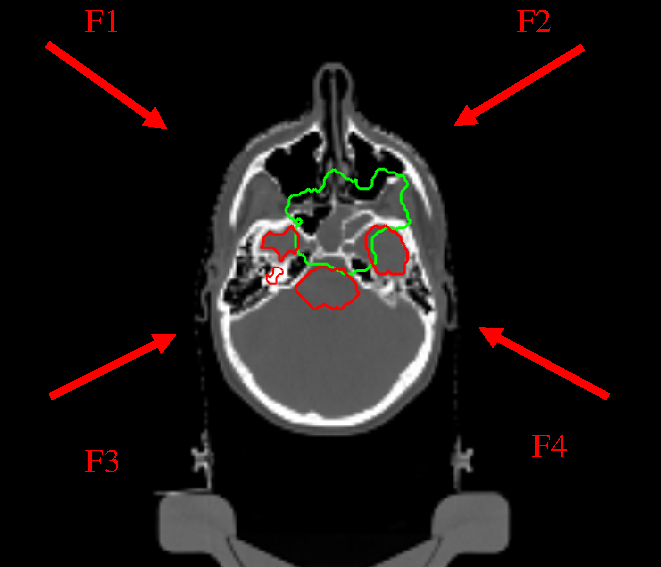}
                \caption{Empty nasal cavity}
        \end{subfigure}
        \hspace*{1cm}
        \begin{subfigure}[b]{0.3\textwidth}
                \includegraphics[scale=0.3]{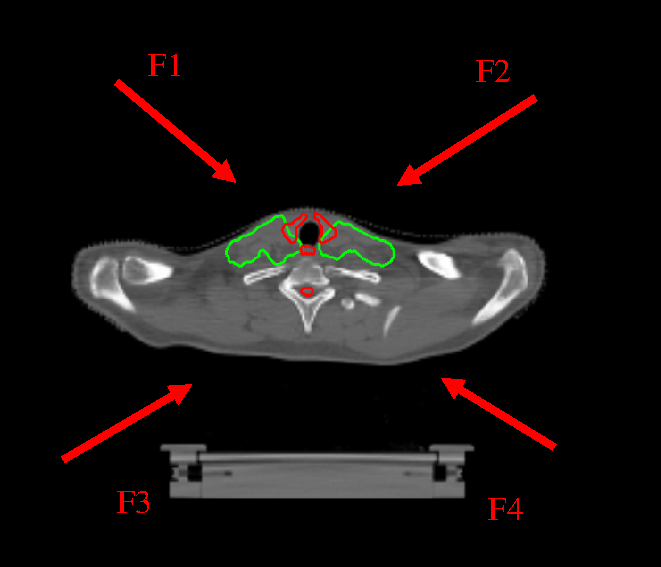}
                \caption{Nasal cavity filled with mucus}
        \end{subfigure}
        \caption{Fields crossing the nasal cavities: axial CT slices with the green contour of the CTV and red contours of some OARs;
                 the four fields F1, F2, F3, F4 are also indicated.}
        \label{head-neck-fields-1}
\end{figure*}

\begin{figure*}[!h]
  \centering
  \includegraphics[width=3.5in]{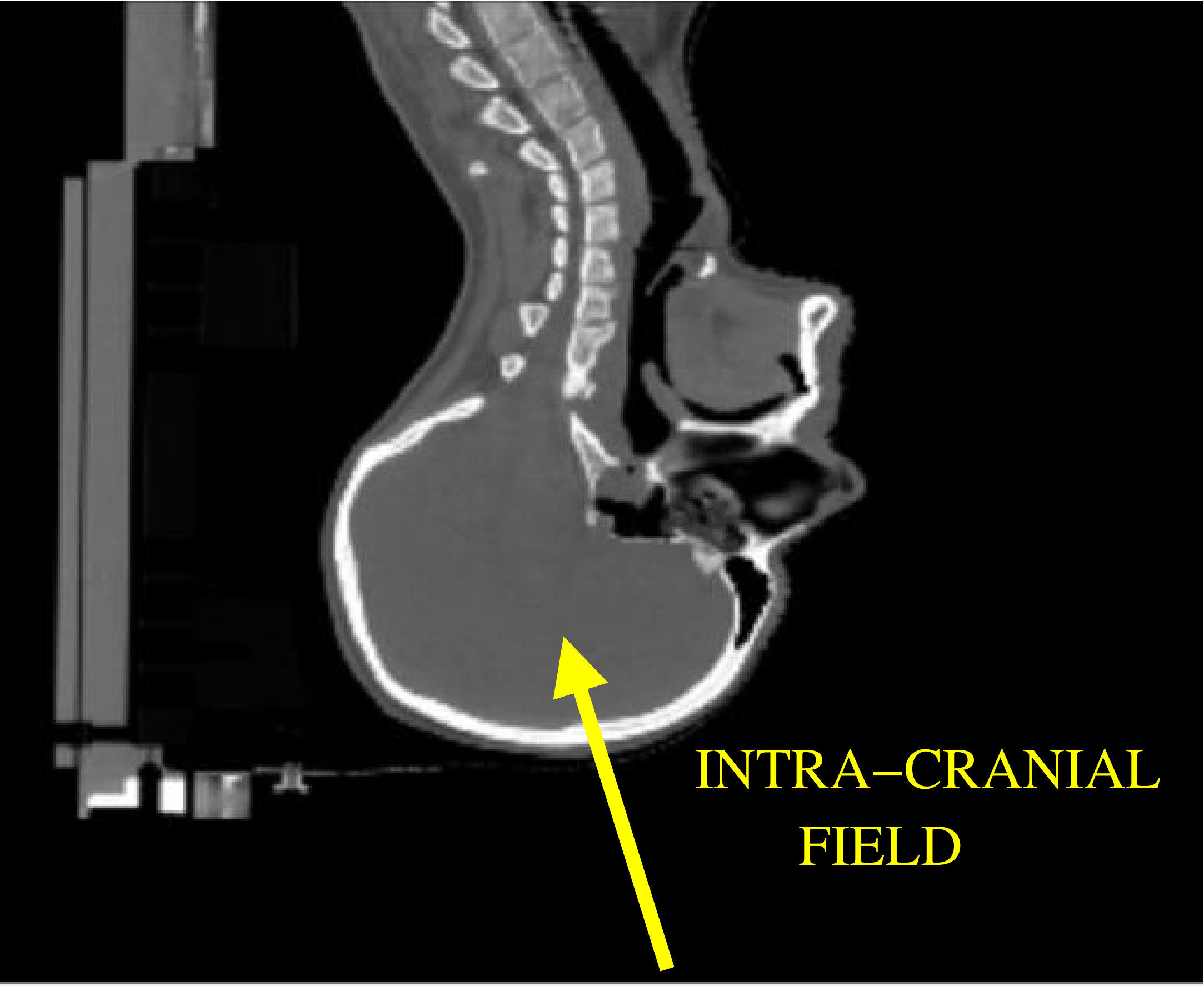}
  \caption{Head and neck tumour: azimuthal CT slice showing the fifth intracranial field added for PLAN-OL-2.}
  \label{head-neck-intracranial-field}
\end{figure*}

One common indication for proton therapy centers is represented by tumours extending both in the head and neck.
As the tumour extends over a relatively big area, the treatment can result rather toxic for many organs at risk.
These cases are usually treated with four fields, as shown in Fig.\ref{head-neck-fields-1}: two coming from the front,
aimed at covering the target volume at the level of the shoulders, two coming from behind, supposed to deliver the dose
in the head part of the tumour. The drawback of this geometry is that, despite the dose constraints,
the front fields irradiate OARs inside the head and, in the same way, the back fields release high dose to the shoulders, without being
crucial for target coverage in that point.
\newline
The OL penalization is, here, exploited to switch off the fields in selected areas and to test a new field geometry.
In PLAN-OL-1, two artificial VOIs were drawn at the level if the shoulders and were assigned with OL = 1,
to switch off, respectively, the posterior fields irradiating that area.
In PLAN-OL-2, two addional artificial VOIs are created at the level of the head and set with OL = 1, to switch off the 
anteriori fields in the part of the patient volume and a fifth intra-cranial field is added to compensate for the target coverage,
as shown in Fig.\ref{head-neck-intracranial-field}.

\section{Results}
The results of the simulations performed with the OL penalization are here shown and compared
to the nominal plans, described in the previous section.

\subsection{Patient with metallic cage}
\begin{figure*}[!h]
        \centering
        \hspace*{-1.5cm}
        \begin{subfigure}[b]{0.3\textwidth}
                \includegraphics[scale=0.3]{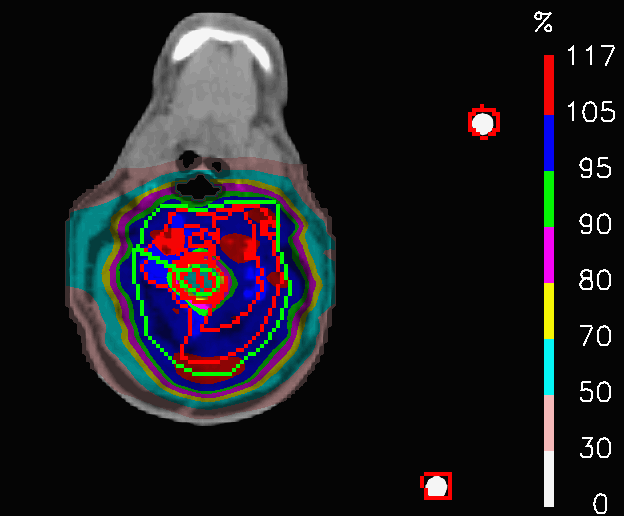}
                \caption{PLAN-NOM}
        \end{subfigure}
        \hspace*{2cm}
        \begin{subfigure}[b]{0.3\textwidth}
                \includegraphics[scale=0.3]{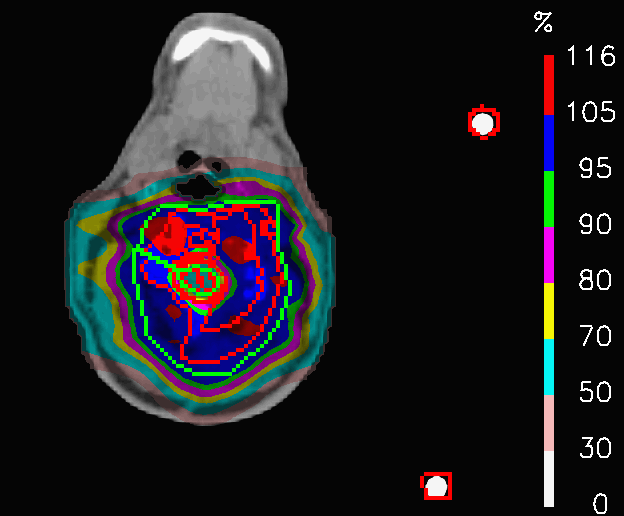}
                \caption{PLAN-OL-1}
        \end{subfigure}
        \caption{Patient with metallic cage: dose distributions for the nominal plan and the one re-computed with OL penalization, keeping
                 the same field configuration.}
        \label{metallic-cage-plan-1}
\end{figure*}

\begin{figure*}[!h]
        \centering
        \hspace*{-3.0cm}
        \begin{subfigure}[b]{0.3\textwidth}
                \includegraphics[scale=0.35]{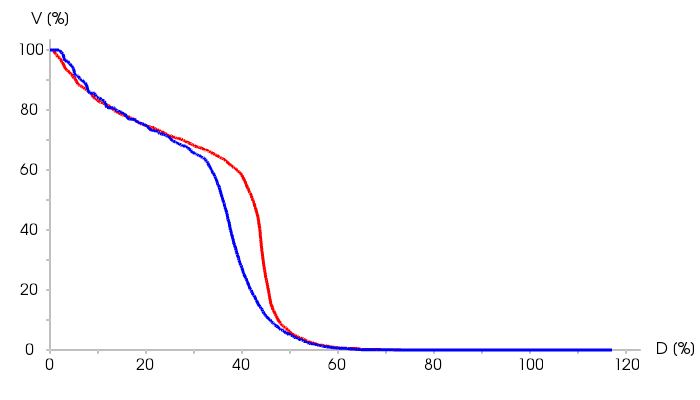}
                \caption{Cumulative DVHs of the left parotis}
        \end{subfigure}
        \hspace*{3.0cm}
        \begin{subfigure}[b]{0.3\textwidth}
                \includegraphics[scale=0.35]{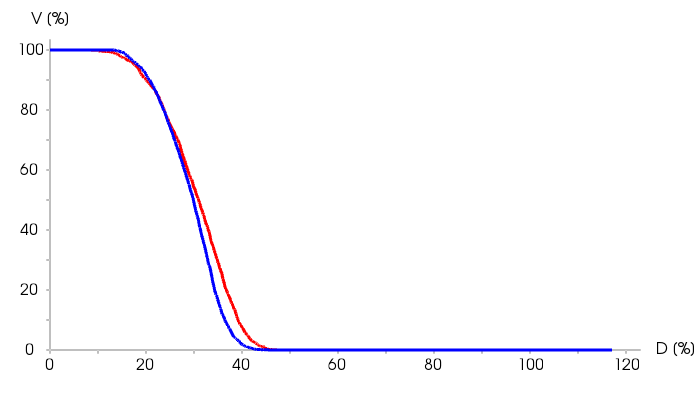}
                \caption{Cumulative DVHs of the right parotis}
        \end{subfigure}
        \caption{Patient with metallic cage: comparison between the cumulative DVHs of PLAN-NOM (red line) and PLAN-OL-1 (blue line) for 
                 the left and right parotis.}
        \label{metallic-cage-parotis-dvh-1}
\end{figure*}

\begin{figure*}[!h]
        \centering
        \hspace*{-1.5cm}
        \begin{subfigure}[b]{0.3\textwidth}
                \includegraphics[scale=0.3]{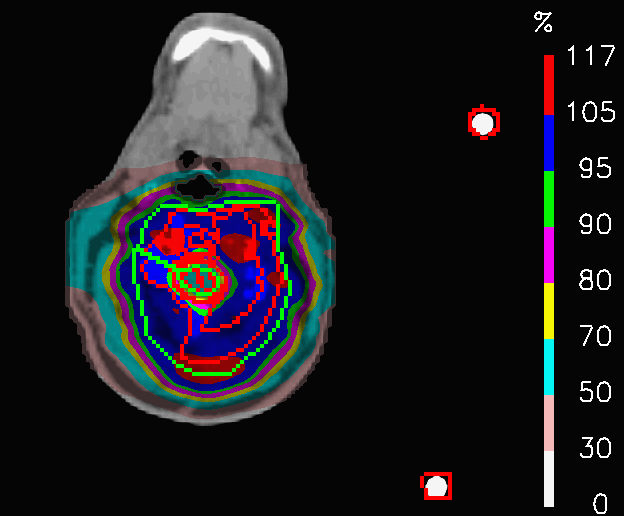}
                \caption{PLAN-NOM}
        \end{subfigure}
        \hspace*{2cm}
        \begin{subfigure}[b]{0.3\textwidth}
                \includegraphics[scale=0.3]{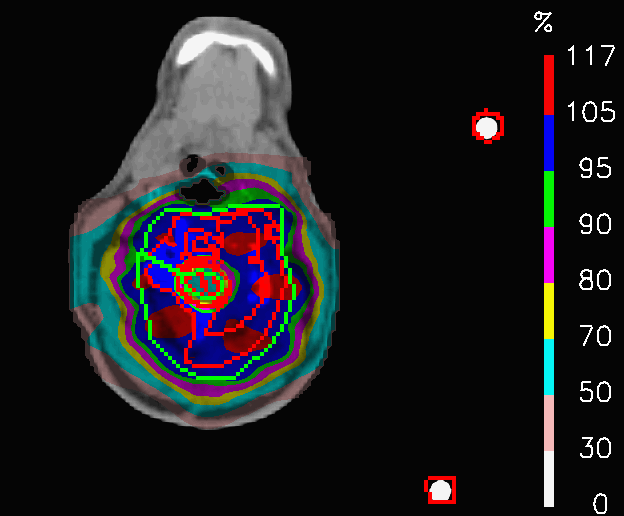}
                \caption{PLAN-OL-1}
        \end{subfigure}
        \caption{Patient with metallic cage: dose distributions for the nominal plan and the one re-computed with OL penalization, choosing
                 new fields more angularly spaced (+10\textdegree) with respect to those used for the nominal plan.}
        \label{metallic-cage-plan-2}
\end{figure*}

\begin{figure*}[!h]
  \centering
  \includegraphics[width=3.5in]{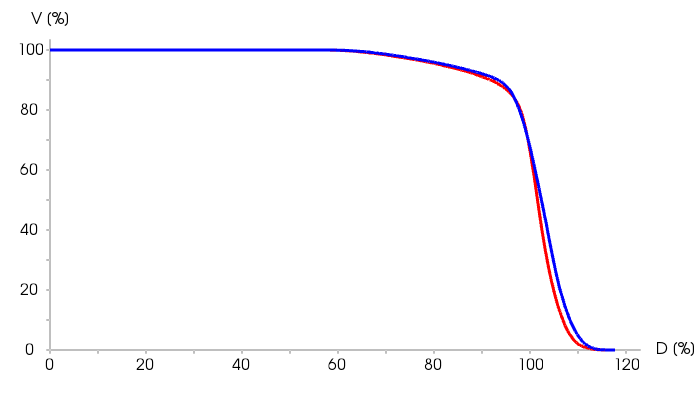}
  \caption{Patient with metallic cage: comparison of the cumulative DVHs of the CTV for PLAN-NOM and PLAN-OL-2.}
  \label{metallic-cage-parotis-ctv-2}
\end{figure*}

Both plans with OL penalization have been re-calculated by dropping the dose constraints on the
safety margins drawn around the metallic rods and by setting the VOIs of the actual rod contours
with OL = 1. In this way, all pencil beams crossing even a voxel of such strutures are completely
switched off and they cannot be activated in the subsequent optimization.
\newline
In Fig.\ref{metallic-cage-plan-1}, the dose distributions of PLAN-NOM and PLAN-OL-1 are shown, but it
is from tables \ref{table-metallic-cage-1} that we get an insight regarding the main differences of the 
plans. In PLAN-OL-1, the dose coverage of the clinical target volume (CTV) improves significantly leading
to an increase of +5.4\% of the volume reached by 100\% of the prescribed dose ($\mathbf{V_{100}}$).
Moreover, left and right parotis, the myelon and the brainstem are characterized by a substantial decrease in both
the maximum and the mean dose ($\mathbf{D_{max}}$, $\mathbf{D_{mean}}$) in PLAN-OL-1. The improved sparing of the parotis is 
well represented by the comparison of the cumulative DVHs in Fig.\ref{metallic-cage-parotis-dvh-1}, where the red line refers
to PLAN-NOM and the blue one to PLAN-OL-1.
\newline
PLAN-OL-2 was conceived with more angularly spaced fields than the original F1, F2, F3, F4; this configuration would
have been not feasible without the OL penalization, since the tough constraints of the rod safety margins would have 
prevented a homogeneous coverage of the target volume, leading necessarily to underdosed areas in the periphery of the CTV
and high dose peaks in the middle. Fig.\ref{metallic-cage-plan-2} shows the dose distributions of PLAN-NOM and PLAN-OL-2,
whereas results are summarized in tables \ref{table-metallic-cage-2}. Despite the fact the increase of $\mathbf{V_{100}}$
is less accentuated in PLAN-OL-2 than in PLAN-OL-1, the improved target coverage is visible in the comparison of the 
cumulative DVHs of the CTV in Fig.\ref{metallic-cage-parotis-ctv-2}, whereas no difference was noticeable in the cumulative
DVH of PLAN-OL-1 compared to PLAN-NOM. This new field configuration provides a remarkable sparing of the left and right parotis
and the brainstem at the cost of decreasing the improvement concerning the myelon.

\begin{figure*}[!h]
        \centering
        \hspace*{-3.0cm}
        \begin{subfigure}[b]{0.3\textwidth}
                \includegraphics[scale=0.35]{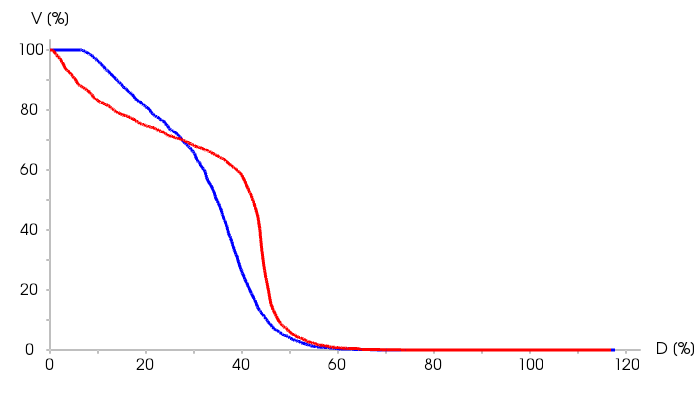}
                \caption{Cumulative DVHs of the left parotis}
        \end{subfigure}
        \hspace*{3.0cm}
        \begin{subfigure}[b]{0.3\textwidth}
                \includegraphics[scale=0.35]{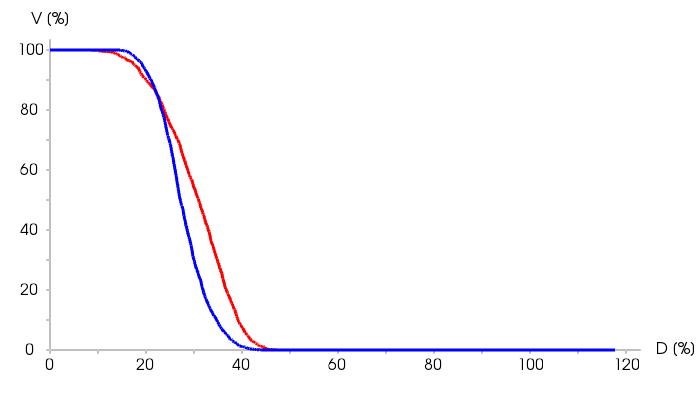}
                \caption{Cumulative DVHs of the right parotis}
        \end{subfigure}
        \caption{Patient with metallic cage: comparison between the cumulative DVHs of PLAN-NOM (red line) and PLAN-OL-2 (blue line) for 
                 the left and right parotis.}
        \label{metallic-cage-parotis-dvh-2}
\end{figure*}

\begin{table*}[!b]\centering
  \ra{1.3}
  \hspace*{-1cm}
  \begin{tabular}{@{}rrrrcrrrcrrr@{}}\toprule
    & \phantom{abc} & CTV & \phantom{abc} & PTV \\ \midrule
    $\mathbf{D_{mean}}$     && +0.7\% && +1.2\% \\ \midrule
    $\mathbf{V_{100}}$     && +5.4\% && +7.2\% \\
    \bottomrule
  \end{tabular}
  \hspace*{1.0cm}
  \begin{tabular}{@{}rrrrcrrrcrrr@{}}\toprule
    & \phantom{abc} & $\mathbf{D_{max}}$ & \phantom{abc} & $\mathbf{D_{mean}}$ \\ \midrule
    Right parotis     && -1.7\% && -1.1\% \\ \midrule
    Left parotis      && -3.3\% && -2.7\% \\ \midrule
    Myelon center     && -2.5\% && -0.4\% \\ \midrule
    Brainstem center  && -1.8\% && -1.7\% \\ 
    \bottomrule
  \end{tabular}\vskip 5 pt
  \caption{Patient with metallica cage: tables summarizing the differences between PLAN-OL-1 and PLAN-NOM concerning the dose coverage
           of the target (left table) and OARs (right table).}
  \label{table-metallic-cage-1}
\end{table*}

\begin{table*}[!t]\centering
  \ra{1.3}
  \hspace*{-1cm}
  \begin{tabular}{@{}rrrrcrrrcrrr@{}}\toprule
    & \phantom{abc} & CTV & \phantom{abc} & PTV \\ \midrule
    $\mathbf{D_{mean}}$     && +0.9\% && +1.2\% \\ \midrule
    $\mathbf{V_{100}}$     && +1.5\% && +3.4\% \\
    \bottomrule
  \end{tabular}
  \hspace*{1.0cm}
  \begin{tabular}{@{}rrrrcrrrcrrr@{}}\toprule
    & \phantom{abc} & $\mathbf{D_{max}}$ & \phantom{abc} & $\mathbf{D_{mean}}$ \\ \midrule
    Right parotis     && -3.4\% && -2.4\% \\ \midrule
    Left parotis      && -6.0\% && -1.4\% \\ \midrule
    Myelon center     && -1.3\% && -0.5\% \\ \midrule
    Brainstem center  && -2.1\% && -1.8\% \\
    \bottomrule
  \end{tabular}\vskip 5 pt
  \caption{Patient with metallic cage: tables summarizing the differences between PLAN-OL-2 and PLAN-NOM concerning the dose coverage
           of the target (left table) and OARs (right table).}
  \label{table-metallic-cage-2}
\end{table*}

\subsection{Fields crossing the nasal cavities}
\begin{figure*}[!h]
        \centering
        \hspace*{-1.5cm}
        \begin{subfigure}[b]{0.3\textwidth}
                \includegraphics[scale=0.3]{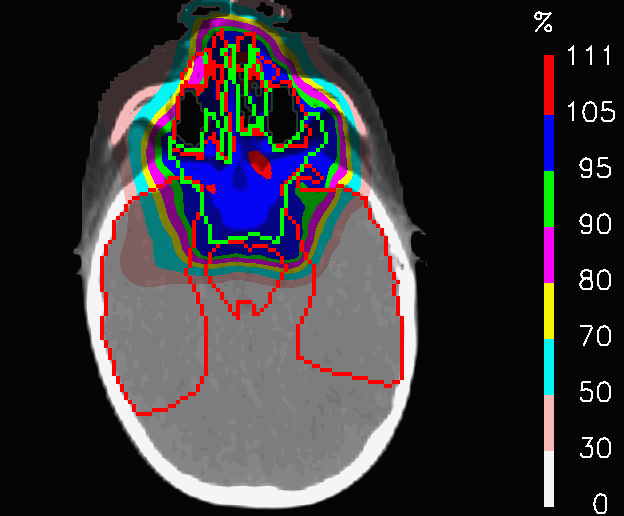}
                \caption{PLAN-NOM}
        \end{subfigure}
        \hspace*{2cm}
        \begin{subfigure}[b]{0.3\textwidth}
                \includegraphics[scale=0.3]{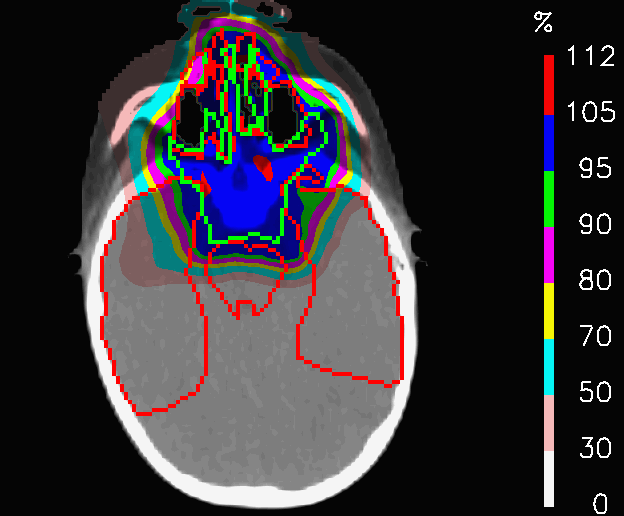}
                \caption{PLAN-OL}
        \end{subfigure}
        \caption{Fields crossing the nasal cavities: dose distributions for the nominal plan and the one re-computed with OL penalization
                 for the nasal cavity case.}
        \label{nasal-cavity-plan}
\end{figure*}

To re-calculate the plan on the original CT with OL penalization, the VOI of the nasal cavity was drawn and set
with an OL = 0.5, meaning that pencil beams were decreased in fluence proportionally to their path length inside the VOI,
according to formula (\ref{smooth-penalization}). In the nominal plan, no dose constraints or other countermeasures were
taken into consideration to deal with the potential density changes of the nasal cavities, since they lie close to the target
volume and any dose constraint would severely affect the target coverage.
\newline
Fig.\ref{nasal-cavity-plan} shows the dose distribution of PLAN-NOM and PLAN-OL, that were computed on the original CT.
The plans result to be identical up to differences of 0.2\%,
as far as concerns both the target coverage and the sparing of the OARs.
\newline
When PLAN-NOM and PLAN-OL are re-computed on the extreme scenario CTs of Fig.\ref{nasal-cavities} and the difference 
between each of these extreme-case plan and the corresponding reference ones are considered,
interesting results emerge. Fig.\ref{nasal-cavity-plan} shows PLAN-NOM-H and PLAN-OL-H, whereas tables \ref{table-nasal-cavity}
highlights the fact that in both scenarios the target coverage is more remarkably more robust for PLAN-OL than for PLAN-NOM,
since $\mathbf{V_{100}}$ improves by 3.1\% in the high-density nose cavities case and by +5.3\% in the other one. 

\begin{figure*}[!h]
        \centering
        \hspace*{-1.5cm}
        \begin{subfigure}[b]{0.3\textwidth}
                \includegraphics[scale=0.3]{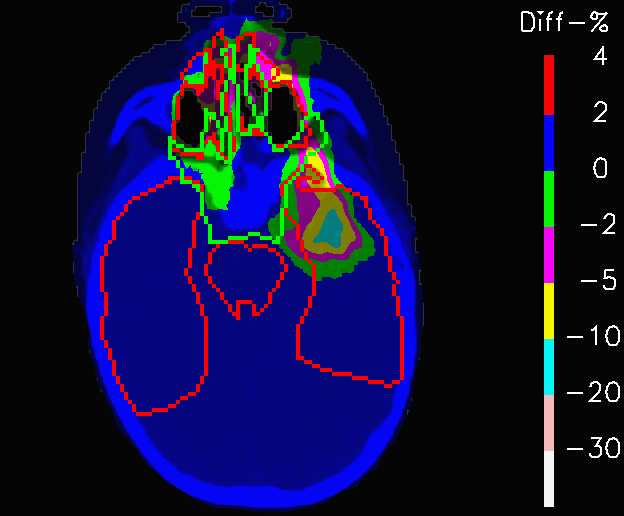}
                \caption{PLAN-NOM-H}
        \end{subfigure}
        \hspace*{2cm}
        \begin{subfigure}[b]{0.3\textwidth}
                \includegraphics[scale=0.3]{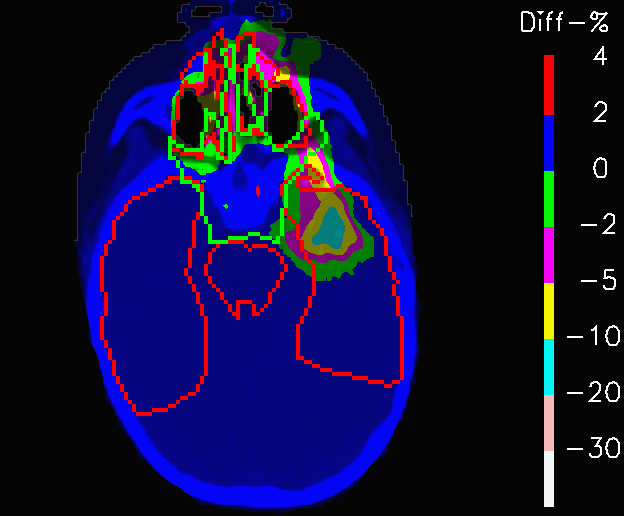}
                \caption{PLAN-OL-H}
        \end{subfigure}
        \caption{Fields crossing the nasal cavities: difference between the dose distribution re-calculated on the
                 CT with nasal cavities filled with mucus and the one computed on the original CT, 
                 for the nominal plan and the one re-computed with OL penalization.}
        \label{nasal-cavity-plan-high-diff}
\end{figure*}

\begin{table*}[!t]\centering
  \ra{1.3}
  \hspace*{-1cm}
  \begin{tabular}{@{}rrrrcrrrcrrr@{}}\toprule
    & \phantom{abc} & CTV & \phantom{abc} \\ \midrule
    $\mathbf{D_{mean}}$     && +0.4\% \\ \midrule
    $\mathbf{V_{100}}$     && +3.1\% \\
    \bottomrule
  \end{tabular}
  \hspace*{1.0cm}
  \begin{tabular}{@{}rrrrcrrrcrrr@{}}\toprule
    & \phantom{abc} & CTV & \phantom{abc} \\ \midrule
    $\mathbf{D_{mean}}$     && +0.5\% \\ \midrule    
    $\mathbf{V_{100}}$     && +5.3\% \\
    \bottomrule
  \end{tabular}
  \caption{Fields crossing the nasal cavities: tables summarizing the differences between PLAN-OL-H and PLAN-NOM-H (on the left) 
           and PLAN-OL-L  and PLAN-NOM-L (on the right) concerning the dose coverage
           of the target.}
  \label{table-nasal-cavity}
\end{table*}

\subsection{Head and neck tumour}
\begin{figure*}[!t]
        \centering
        \hspace*{-1.5cm}
        \begin{subfigure}[b]{0.3\textwidth}
                \includegraphics[scale=0.3]{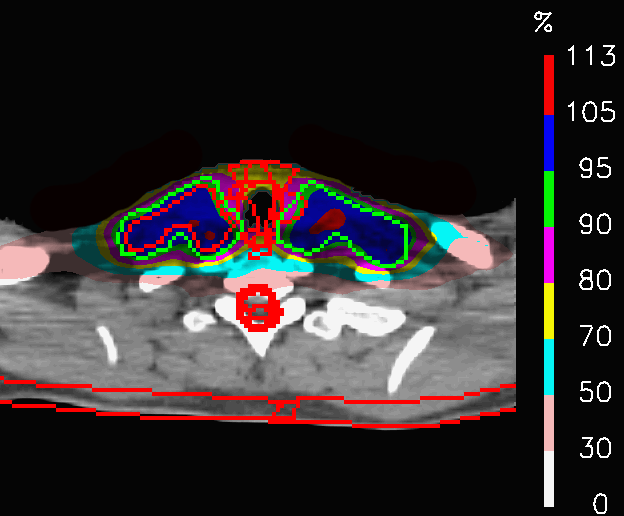}
                \caption{PLAN-NOM}
        \end{subfigure}
        \hspace*{2cm}
        \begin{subfigure}[b]{0.3\textwidth}
                \includegraphics[scale=0.3]{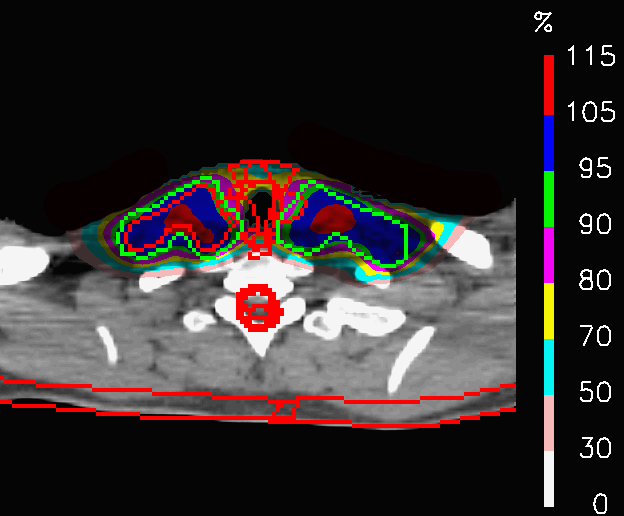}
                \caption{PLAN-OL-1}
        \end{subfigure}
        \caption{Head and neck tumour: dose distributions of PLAN-NOM and PLAN-OL-1 at the level of the shoulders.}
        \label{head-neck-plan-1}
\end{figure*}

\begin{figure*}[!t]
        \centering
        \hspace*{-1.5cm}
        \begin{subfigure}[b]{0.3\textwidth}
                \includegraphics[scale=0.3]{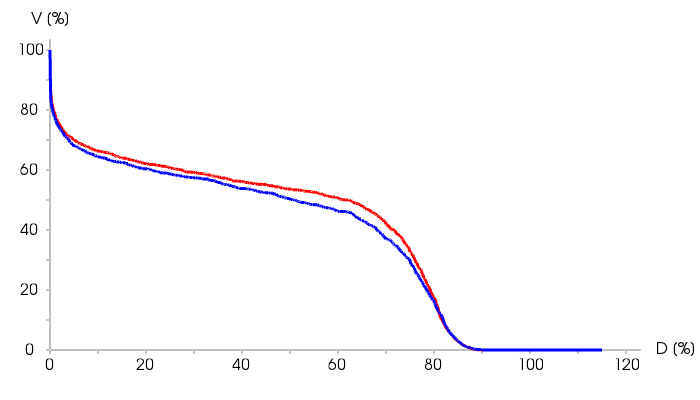}
                \caption{Esophagus}
        \end{subfigure}
        \hspace*{2cm}
        \begin{subfigure}[b]{0.3\textwidth}
                \includegraphics[scale=0.3]{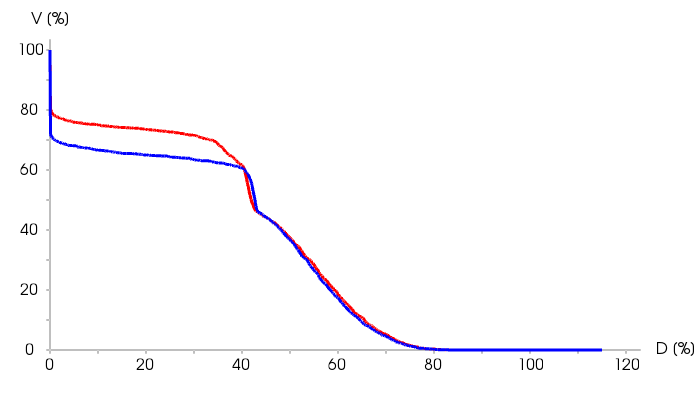}
                \caption{Spinal cord}
        \end{subfigure}
        \caption{Head and neck tumour: comparison between the cumulative DVHs of PLAN-NOM (red line) and PLAN-OL-1 (blue line)
                 for the esophagus and the spinal cord.}
        \label{head-neck-dvh-1}
\end{figure*}

\begin{figure*}[!t]
        \centering
        \hspace*{-1.5cm}
        \begin{subfigure}[b]{0.3\textwidth}
                \includegraphics[scale=0.3]{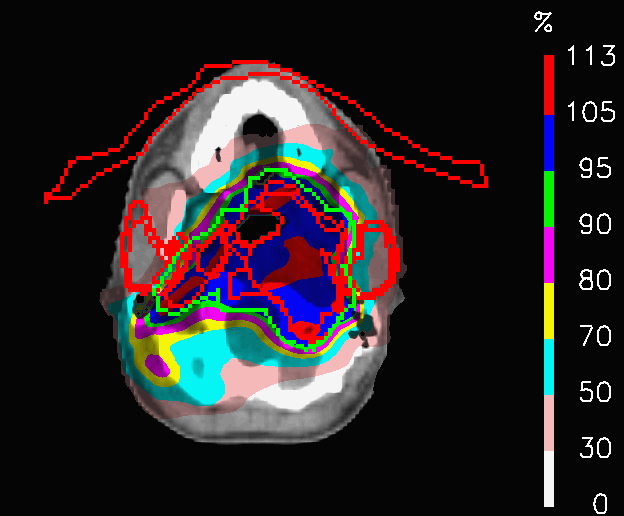}
                \caption{Esophagus}
        \end{subfigure}
        \hspace*{2cm}
        \begin{subfigure}[b]{0.3\textwidth}
                \includegraphics[scale=0.3]{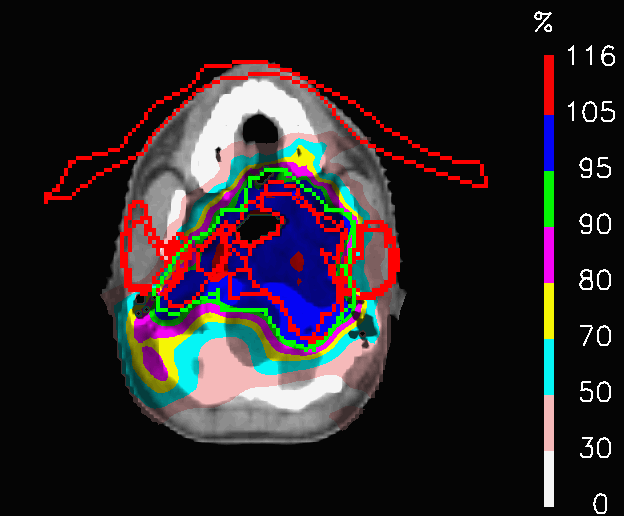}
                \caption{Spinal cord}
        \end{subfigure}
        \caption{Head and neck tumour: dose distributions of PLAN-NOM and PLAN-OL-2 at the level of the head.}
        \label{head-neck-plan-2}
\end{figure*}

\begin{figure*}[!t]
        \centering
        \hspace*{-1.5cm}
        \begin{subfigure}[b]{0.3\textwidth}
                \includegraphics[scale=0.3]{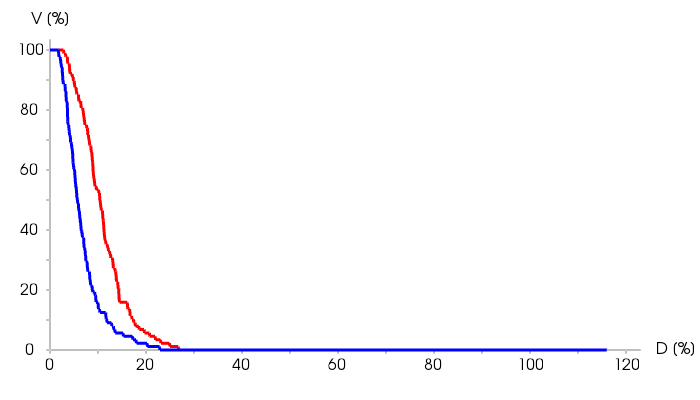}
                \caption{Left lense}
        \end{subfigure}
        \hspace*{2cm}
        \begin{subfigure}[b]{0.3\textwidth}
                \includegraphics[scale=0.3]{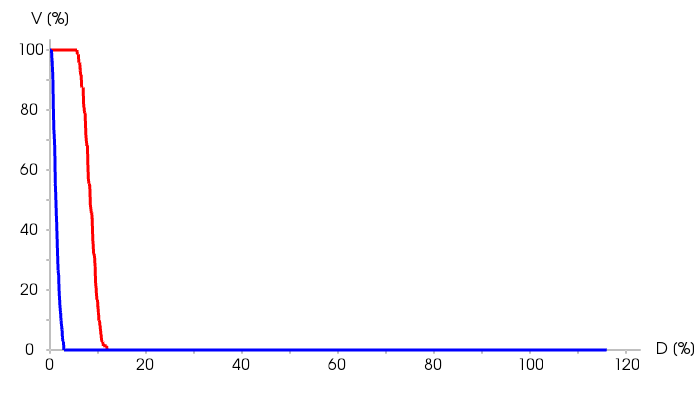}
                \caption{Right lense}
        \end{subfigure}\\\vskip 5 pt
        \begin{subfigure}[b]{0.3\textwidth}
                \includegraphics[scale=0.3]{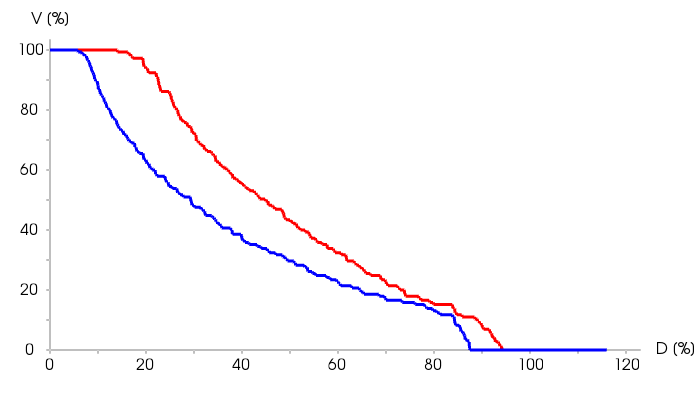}
                \caption{Right optical nerve}
        \end{subfigure}      
        \caption{Head and neck tumour: comparison between the cumulative DVHs of PLAN-NOM (red line) and PLAN-OL-2 (blue line)
                 for the left and right lense and the right optical nerve.}
        \label{head-neck-dvh-2}
\end{figure*}

In this clinical case, the OL penalization is aimed at optimizing the same fields of the nominal plan
(PLAN-OL-1) and at testing a new treatment configuration (PLAN-OL-2).
\newline
By comparing the dose distributions of PLAN-NOM and PLAN-OL-1 at the level of the shoulders in Fig.\ref{head-neck-plan-1}, it is
noticeable how the dose in PLAN-OL-1 results better confined inside the CTV. This fact determines a remarkable improvement 
in the sparing of the esophagus and the spinal cord, as reported in Table \ref{table-head-neck-1} and in Fig.\ref{head-neck-dvh-1}, 
where the comparison between the cumulative DVHs of the two plans are shown. No relevant changes concern the coverage of the target
volume, since $\mathbf{D_{mean}}$ differs by 0.3\% from the nominal plan. These results clearly show that switching off the posterior
fields at the level of the shoulders allows to achieve a better sparing of some OARs, while keeping the same irradiation level for the
tumour.
\newline
In PLAN-OL-2, a different field geometry has been investigated: the posterior fields are switched off at the level of the shoulders,
the anterior ones at the level of the head and a fifth intracranial field is added (Fig.\ref{head-neck-intracranial-field}). The goal
was to achieve a better preservation of the OARs inside the head. The dose distributions (Fig.\ref{head-neck-plan-2}) share
a very similar coverage of the target volume, but they are characterized by substantial differences
in the dose delivered to the OARs, as shown by Table \ref{table-head-neck-2}. In fact, the huge improvements in the sparing of the chiasm,
right lens (Fig.\ref{head-neck-dvh-2}a), left lens (\ref{head-neck-dvh-2}b), left inner ear and the spinal cord (considering in first place the decrease of $\mathbf{D_{max}}$) are followed 
by a significant increase of the peak dose inside the thyroid and the mean dose to the brainstem. In this case, PLAN-OL-2 is not necessarily
superior to PLAN-NOM, but it is undeniable that the OL penalization has allowed the planner to test a scenario, that the simple enforcement
of the dose constraints would not have provided.

\begin{table*}[!t]\centering
  \ra{1.3}
  \hspace*{1.0cm}
  \begin{tabular}{@{}rrrrcrrrcrrr@{}}\toprule
    & \phantom{abc} & $\mathbf{D_{max}}$ & \phantom{abc} & $\mathbf{D_{mean}}$ \\ \midrule
    Right inner ear   && -0.3\% && -1.2\% \\ \midrule
    Esophagus         &&   /    && -2.2\% \\ \midrule
    Spinal cord       && -0.2\% && -3.2\% \\ \midrule
    Brainstem         && -1.0\% && -0.6\% \\
    \bottomrule
  \end{tabular}\vskip 5 pt
  \caption{Head and neck tumour: tables summarizing the differences between PLAN-OL-1 and PLAN-NOM concerning the 
           the sparing of the OARs.}
  \label{table-head-neck-1}
\end{table*}

\begin{table*}[!t]\centering
  \ra{1.3}
  \hspace*{1.0cm}
  \begin{tabular}{@{}rrrrcrrrcrrr@{}}\toprule
    & \phantom{abc} & $\mathbf{D_{max}}$ & \phantom{abc} & $\mathbf{D_{mean}}$ \\ \midrule
    Right lens          && -9.0\% && -7.1\% \\ \midrule
    Left lens           && -3.9\% && -4.1\% \\ \midrule
    Left inner ear      && -2.6\% && +0.3\% \\ \midrule
    Left parotis gland  && -0.2\% && -3.9\% \\ \midrule
    Thyroid             && +2.7\% && +4.9\% \\ \midrule
    Spinal cord         && -2.0\% && +2.0\% \\ \midrule
    Brainstem           && -1.9\% && +9.5\% \\ \midrule
    Chiasm              && -13.9\% && -4.3\% \\
    \bottomrule
  \end{tabular}\vskip 5 pt
  \caption{Head and neck tumour: tables summarizing the differences between PLAN-OL-2 and PLAN-NOM concerning the
           the sparing of the OARs.}
  \label{table-head-neck-2}
\end{table*}

\clearpage\newpage


\section{Discussion}
In this work, we have introduced the concept of \emph{opacity} as a quality assigned to VOIs  and aimed at steering the outcome of the 
IMPT planning by changing the starting condition of the optimization. In particular, the opacity level (OL) of a VOI is a weight
ranging in [0,1] and is used to penalize the initial fluence of the pencil beams crossing the voxels of such structure: 
when OL = 0, the VOI is transparent and no penalization is performed; when  0 $<$ OL $<$ 1, the pencil beams are decreased in fluence proportionally to
their path length inside the VOI; when OL = 1, they are totally switched off. 
\newline
We have shown how this simple additional feature can lead to dose distributions of higher quality and robustness within the framework
the IMPT treatment planning. Three different clinical cases were taken into consideration to show the potential advantage of adopting the 
OL penalization in the pre-optimization stage: a patient with a post-surgery metallic cage, a treatment where fields cross the nasal cavities and
the case of a complex head and neck tumour.
\newline
Concerning the first clinical case, the opacity tool has shown to perfectly tackle the presence of metallic structures:
the safety margins and hard dose constraint on the metal are replaced with the actual contour of the structure and set with OL = 1.
This approach has lead to a plan, where both the coverage of the target volume and the sparing of some OARs have improved. The strategy to handle 
the metallic cage can be re-proposed for golden teeth or any other prothesis, whose composition and, therefore, calibration curve are not known.
\newline
As far as regards the second clinical case, the OL penalization has provided a more robust dose distribution in relation to the possible density
changes inside the nasal cavities. Since the the plan does not change when OL is set to a specific value in (0,1), we obtain a simple way to guarantee
the stability of the treatment. This approach may be applied for any other anatomical cavity, placed close to the tumour and that can undergo substantial
density variations within the treatment frametime.
\newline
For the case of the head and neck tumour, we have shown how the opacity allows to optimize the usage of each single field, especially when dealing
with a tumour mass of complex shape and considerable extension. We have also pointed out how the opacity can easily shape the fields in order to test
treatment scenario, that would not be possible to achieve only through dose constraints.
\newline
In conclusion, the opacity is a leverage that be used by the planner to mould the IMPT dose distributions, when seeking for the best tradeoff 
between target coverage and sparing of the OARs. It is no meant to replace the dose constraints, but rather to be played as an additional card
in the IMPT treatment planning.


\bibliographystyle{IEEEtran}
\bibliography{opacity}

\begin{thebibliography}{1}
\providecommand{\url}[1]{#1}
\csname url@samestyle\endcsname
\providecommand{\newblock}{\relax}
\providecommand{\bibinfo}[2]{#2}
\providecommand{\BIBentrySTDinterwordspacing}{\spaceskip=0pt\relax}
\providecommand{\BIBentryALTinterwordstretchfactor}{4}
\providecommand{\BIBentryALTinterwordspacing}{\spaceskip=\fontdimen2\font plus
\BIBentryALTinterwordstretchfactor\fontdimen3\font minus
  \fontdimen4\font\relax}
\providecommand{\BIBforeignlanguage}[2]{{%
\expandafter\ifx\csname l@#1\endcsname\relax
\typeout{** WARNING: IEEEtran.bst: No hyphenation pattern has been}%
\typeout{** loaded for the language `#1'. Using the pattern for}%
\typeout{** the default language instead.}%
\else
\language=\csname l@#1\endcsname
\fi
#2}}
\providecommand{\BIBdecl}{\relax}
\BIBdecl

\bibitem{Pedroni1995}
\BIBentryALTinterwordspacing
E.~Pedroni, R.~Bacher, H.~Blattmann, T.~B\"{o}hringer, A.~Coray, A.~Lomax,
  S.~Lin, G.~Munkel, S.~Scheib, U.~Schneider, and A.~Tourovsky, ``The 200-{MeV}
  proton therapy project at the paul scherrer institute: Conceptual design and
  practical realization,'' \emph{Medical Physics}, vol.~22, no.~1, pp. 37--53,
  jan 1995. [Online]. Available: \url{https://doi.org/10.1118%2F1.597522}
\BIBentrySTDinterwordspacing

\bibitem{Scheib1993}
\BIBentryALTinterwordspacing
S.~Scheib, ``Spot-scanning mit protonen,'' 1993. [Online]. Available:
  \url{http://dx.doi.org/10.3929/ethz-a-000945359}
\BIBentrySTDinterwordspacing

\bibitem{Lomax1999}
\BIBentryALTinterwordspacing
A.~Lomax, ``Intensity modulation methods for proton radiotherapy,''
  \emph{Physics Medicine Biology}, vol.~44, pp. 185--205, 1999. [Online].
  Available:
  \url{http://iopscience.iop.org/article/10.1088/0031-9155/44/1/014/pdf}
\BIBentrySTDinterwordspacing

\bibitem{Albertini2011}
\BIBentryALTinterwordspacing
F.~Albertini, ``Planning and optimizing treatment plans for actively scanned
  proton therapy: evaluating and estimating the effect of uncertainties,''
  2011. [Online]. Available: \url{http://dx.doi.org/10.3929/ethz-a-006576001}
\BIBentrySTDinterwordspacing

\bibitem{Bortfeld1996}
\BIBentryALTinterwordspacing
T.~Bortfeld and W.~Schlegel, ``An analytical approximation of depth - dose
  distributions for therapeutic proton beams,'' \emph{Physics in Medicine and
  Biology}, vol.~41, no.~8, p. 1331, 1996. [Online]. Available:
  \url{http://stacks.iop.org/0031-9155/41/i=8/a=006}
\BIBentrySTDinterwordspacing

\bibitem{Jacobs1998}
\BIBentryALTinterwordspacing
B.~D. S. M. C. I.~L. Filip~Jacobs, Erik~Sundermann, ``A fast algorithm to
  calculate the exact radiological path through a pixel or voxel space,''
  \emph{CIT. Journal of computing and information technology}, vol.~6, no.~1,
  pp. 89--94, 1998. [Online]. Available:
  \url{users.elis.ugent.be/~brdsutte/research/publications/1998JCITjacobs.pdf}
\BIBentrySTDinterwordspacing

\end{thebibliography}

\end{document}